# The Effects of EGF-Receptor Density on Multiscale Tumor Growth Patterns

**Chaitanya A. Athale [1] and Thomas S. Deisboeck [1,*]**

[1] Complex Biosystems Modeling Laboratory, Harvard-MIT (HST) Athinoula A. Martinos Center for Biomedical Imaging, Massachusetts General Hospital, Charlestown, MA 02129.

***Corresponding Author:**

Thomas S. Deisboeck, M.D.
Complex Biosystems Modeling Laboratory
Harvard-MIT (HST) Athinoula A. Martinos Center for Biomedical Imaging
Massachusetts General Hospital-East, 2301
Bldg. 149, 13th Street
Charlestown, MA 02129
Tel: 617-724-1845
Fax: 617-726-5079
Email: deisboec@helix.mgh.harvard.edu

## ABSTRACT

We studied the effects of epidermal growth factor receptor (EGFR) density on tumor growth dynamics, both on the sub- and the multi-cellular level using our previously developed model. This algorithm simulates the growth of a brain tumor using a multi-scale two-dimensional agent-based approach with an integrated transforming growth factor α (TGFα) induced EGFR-gene-protein interaction network. The results confirm that increasing cell receptor density correlates with an acceleration of the tumor system's spatio-temporal expansion dynamics. This multicellular behavior cannot be explained solely on the basis of spatial sub-cellular dynamics, which remain qualitatively similar amongst the three glioma cell lines investigated here *in silico*. Rather, we find that cells with higher EGFR density show an early increase in the phenotypic switching activity between proliferative and migratory traits, linked to a higher level of initial auto-stimulation by the PLCγ-mediated TGFα-EGFR autocrine network. This indicates a more active protein level interaction in these chemotactically acting tumor systems and supports the role of post-translational regulation for the implemented EGFR pathway. Implications of these results for experimental cancer research are discussed.

## 1. INTRODUCTION

The epidermal growth factor receptor (EGFR) has been widely implicated in tumorigenesis in the brain (Diedrich et al., 1995). Early clinical studies have found that up to four in ten specimen of highly-malignant brain tumors or gliomas overexpress EGFR due to gene amplification (Libermann et al., 1985a, 1985b). Furthermore, this EGFR amplification correlates with rapid glioma growth and invasion (Lund-Johansen et al., 1990; Lund-Johansen et al., 1992; Schlegel et al., 1994; Westermark et al., 1982) thus rendering the EGFR signaling pathway of importance for a better understanding of tumor growth dynamics.

Detailed steady state modeling of EGFR dynamics has already investigated the binding, internalization and degradation of ligands (Wiley and Cunningham, 1981). Later work has addressed aspects like the role of endocytosis (Starbuck and Lauffenburger, 1992), differential feedback in response to different ligands (Brightman and Fell, 2000) and the robustness of an extended EGFR network to initial values (Schoeberl et al., 2002). Furthermore, spatial 2D models of EGFR signaling have also examined pattern generation with multiple (Shvartsman et al., 2002) and single cells (Maly et al., 2004). Finally, a model of TGFα EGFR signaling (Owen and Sherratt, 1998) has already shown to be capable of pattern formation in a multi-cellular environment (Owen et al., 2000). Taken together, computational modeling and quantitative experimentation of the EGFR pathway have therefore addressed reaction kinetics as well as the effect on cell-behavior like proliferation and migration (reviewed in Lauffenburger and Linderman, 1996).

Giese et al. (1996) first proposed that a *dichotomy* exists between these phenotypes of migration and proliferation in growing tumors, i.e. that cells either replicate or migrate, yet do not display both traits at the same time. Wells (1999) proposed that the EGFR pathway and its downstream elements like Phospholipase Cγ (PLCγ) could provide the molecular mechanism for such *dichotomy*. In fact, experimental evidence has accumulated for a prominent role of Phospholipase Cγ (PLCγ) in mediating migratory and proliferative responses on EGFR stimulation (Chen et al., 1994, 1996). Recent experimental data suggested a correlation between rate of PLCγ concentration changes and the migratory and proliferative phenotypes (Dittmar et al., 2002). We have employed this mechanism in our previously developed agent-based 'microscopic-macroscopic' brain tumor model (Mansury et al., 2002; Mansury and Deisboeck, 2003, 2004a, 2004b). Further incorporating an EGFR molecular gene-protein interaction network and spatial distribution module, we were able to demonstrate that one can indeed simulate the dichotomy between cell migration and proliferation based on an intracellular EGFR-dependent gene-protein decision network (Athale et al., in press).

In order to better understand the interaction between molecular dynamics at the protein level and multicellular tumor growth patterns, specifically in the context of EGFR signaling, we again employ this multiscale cancer model. Here, we, however, vary the expression level of EGFR protein and examine its impact on multicellular tumor dynamics, cellular phenotype and single cell molecular profiles. We find the overall spatio-temporal expansion to be faster in tumors with higher EGFR densities. The very same cells show an earlier onset of phenotypic switching activity, presumably related to a faster change in average PLCγ concentration and to an earlier increase in extracellular TGFα concentration. The following section describes the specific algorithm employed here.

## 2. MODEL

Our specific agent-based model is described elsewhere in detail (Athale et al., in press). In brief, the algorithm spans several orders of magnitude from the sub-cellular and cell-cell interaction level up to the macroscopic tumor scale. In the following, we will briefly summarize the setup.

### 2.1. Sub-cellular dynamics

Here, the smallest unit of the simulated tumor is the *molecular* entity, which forms a gene-protein interaction network governed by mass balance equations. The molecules are related to the EGFR pathway and reactions are first and second order differential equations in which threshold values determine the behavior of the cell. The mass-balance equation takes the following generic form:

$$\frac{dX_i}{dt} = \alpha \cdot X_i - \beta \cdot X_i, \qquad (1)$$

where $\boldsymbol{X_i}$ is the mass of the $\boldsymbol{i}$-th molecule, $\boldsymbol{\alpha}$ is the rate of synthesis or increase and $\beta$ is its rate of degradation or removal of a molecular species. A schematic of the gene-protein interaction network is displayed in **Fig. 1a** while **Table 1** lists the network's individual molecules.

**Table 1**

The interaction network is implemented in the regions of the nucleus, cytoplasm and cell membrane (**Fig. 1b**). A molecule in one or more of these regions is also distributed in four spatially oriented sub-compartments, i.e. North, South, East and West. The molecular flux between these sub-compartments (**Fig. 1c**) is governed by the following equation:

$$\frac{dX_{i,j}}{dt} = k_{in} \cdot \left[X_{i,j-1} + X_{i,j+1}\right] - 2 \cdot k_{out} \cdot X_{i,j}, \qquad (2)$$

where $\boldsymbol{X_{i,j}}$ is the concentration of the molecule $\boldsymbol{X_i}$ in a given compartment $\boldsymbol{j}$, while $\boldsymbol{X_{i,j-1}}$ refers to the neighboring compartment before $\boldsymbol{X_{i,j}}$ and $\boldsymbol{X_{i,j+1}}$ to the compartment after $\boldsymbol{X_{i,j}}$. Here, $\boldsymbol{j}$ (1 to 4) is the compartment number and $k_{in}$ and $k_{out}$ are the flux rate constants into and out of the compartment, respectively.

**Figure 1a**

**Figure 1b**

**Figure 1c**

## 2.2. Cellular decision making

In our cellular *decision* model a tumor cell can 'choose' either to migrate *or* to proliferate, yet does not perform both phenotypic activities at the same time, hence implementing Giese et al. (1996) aforementioned dichotomy concept. Furthermore, Dittmar et al. (2002) could show in breast cancer cells that PLCγ is activated transiently and to a greater extent during migration and more gradually in the proliferative 'mode'. We therefore adopt a simple threshold, $\sigma_{PLC}$, to decide whether the cell should undergo migration or not. Thus the *potential to migrate* is assessed by a cell evaluating the following function:

$$M_n\left[(PLC\gamma)\right] = \left[\frac{d(PLC\gamma)}{dt}\right]_n, \tag{3}$$

where $d(PLC\gamma)/dt$ is the change in concentration of PLCγ over time ($t$) and $n$ is the cell number. If $M_m > \sigma_{PLC}$ the cell becomes eligible to migrate, otherwise it could proliferate or become quiescent (i.e., it neither proliferates nor migrates, yet remains viable). If the cell decides to migrate, the direction of migration is expressed in terms of a local valuation function for each neighboring lattice point $j$ in the *von Neumann* neighborhood as follows:

$$L_j\left[(PLC\gamma),n_j\right]=\left(1-n_j\right)\cdot\left[(PLC\gamma)\right]_j-\underset{j\leq m}{\arg\max}[(PLC\gamma)]_j\cdot\Psi_{PLC}, \tag{4}$$

where $\boldsymbol{L_j}$ is the value of a grid point in the neighborhood of the cell; the neighboring grid locations and compartments are numbered as $\boldsymbol{j}$ (1 to 4). In a given sub-cellular compartment $j$, ***(PLCγ)*** is the concentration of activated PLCγ, ***n*** is the number of cells at that location in the neighborhood and ***m*** is the total number of compartments (here: ***m***=4). $\Psi_{PLC}\in[0,1]$ denotes the so-called *search-precision* parameter, representing the biological equivalent of a receptor-driven, non-erroneous evaluation of the permissibility of the microenvironment as discussed in detail in our previous work (Athale et al., in press; Mansury and Deisboeck, 2003). For instance, when the search precision parameter $\Psi_{PLC}$ = 0, the cell performs a pure random walk, and when $\Psi_{PLC}$ = 1 the cells never commits 'mistakes' and always migrate fully biased to the most permissive location with the highest level of PLCγ (based on **Eq. 4**).

If the change in concentration of active PLCγ is below the migration-threshold, $\boldsymbol{\sigma_{PLC}}$, yet above a set noise threshold, $\boldsymbol{\sigma_n}$, then the cell evaluates the *proliferative potential* $\boldsymbol{P_{prolif}}$ ≥0. This potential is calculated using **Eq. 5**,

$$P_{prolif}\left[(LR)\right]=(LR)-\sigma_{EGFR}, \tag{5}$$

where ***(LR)*** is the concentration of ligand bound phosphorylated TGFα-EGFR complex in a cell. This function is derived from experimental observations citing cell proliferation in relation to an EGF-receptor threshold (Knauer et al., 1984) and an experimentally validated model, which relates receptor occupancy to percent maximal proliferation (Lauffenburger and Linderman, 1996). Again, if neither the conditions for migration nor proliferation are met, the cell turns quiescent. Cell death is currently not considered in our model here.

### 2.3. Cell-cell communication

Every such cell has a molecular interaction network that includes the protein transforming growth factor alpha, TGFα, which is produced by the cell based on a constant rate of secretion (see also **Fig. 1a**). Previous work has pointed to the fact that TGFα capture is rapid, meaning that most of the hormone is captured by the very same cell producing it (Shvartsman et al., 2001). Therefore, in our simulation it is deposited onto the neighboring lattice grid without any explicit diffusion modeled. It may thus act not only autocrine but also in a paracrine manner, i.e. affecting neighboring cells, as has been demonstrated for breast cancer (de Jong et al., 1998).

### 2.4. Environmental conditions

Finally, the virtual environment is made up of a finite number ***S*** of regular points that are part of a torroidal grid lattice in two-dimensions, where $S = R^2$ and ***R*** is the length of a square section of the grid (***R***=200). Each point of the lattice $\boldsymbol{S_{x,y}}$ (where ***x*** and ***y*** indicate the integer location in Euclidean terms) can be either empty or carry a single cell. The simulation recreates a scenario where a mass of tumor cells ($n_{t=0} = 100$) is initialized in the geometric center of the grid, and a cross-sectioned blood vessel located in the North-East (NE) Quadrant acts as a replenished source of nutrients with glucose diffusing down the concentration gradient, and the edge of the rectangular lattice acts as a perfect sink. The simulation is terminated when the first tumor cell reaches the edge of this nutrient source. This time point is used as a measure to determine the spatio-temporal expansion velocity of the tumor system.

## 3. RESULTS

### 3.1. Simulation methods

The model was implemented in Java (Sun Microsystems, Inc., USA) using the agent-based modeling toolkit Repast (version 2.0; http://repast.sourceforge.net). The mass-balance equations representing the gene-protein network were solved using Euler finite time stepping. Simulation runs were performed on a workstation with dual Intel Xeon 2.3GHz processors, connected via gigabit Ethernet to the central file storage system and running Linux. Three runs with different parameter values took 1 hour, 25 min and 26 seconds to complete.

### 3.2. Multicellular patterns

As a measure of the overall tumor growth rate, we examined the effect of varying the cell surface-bound EGFR protein concentration per cell over three orders of magnitude (i.e., ~$10^4$–$10^6$ EGFR per cell) on the spatio-temporal expansion velocity. As noted before, the simulated tumor growth is halted when the first cell enters the edge of the nutrient source located in the NE Quadrant. The results show that tumor growth accelerates with increasing receptor density (**Fig. 2**) and attains a maximum velocity in the range of $10^6$ receptors per cell. The number of receptors reported in a previous study on *in vitro* invasion of glioma cell lines (Lund-Johansen et al., 1990) is superimposed (on the *X-axis*) for comparison.

**Figure 2**

### 3.3. Sub-cellular localization dynamics of PLCγ

To further investigate a potentially molecular cause of these distinct multicellular patterns which emerge both in time (**Fig. 2**) and space (**Fig. 3a**) we first monitored the polarization of the molecular concentration profiles of PLCγ in the most aggressive migratory cell, i.e. the 1st cell to reach the edge of the cross sectional blood vessel, for each of the three cell lines (**Fig. 3b**). Evidently, the time evolution results (scaled to start with the time point of origin of that cell) show a qualitatively very similar PLCγ polarization pattern, turning from homogeneous to a heterogeneous one, where the highest localization is in the cytosolic compartment facing the nutrient source in the NE quadrant.

**Figure 3a**

**Figure 3b**

### 3.4. Phenotypic switching

Interestingly, when investigating the cell line-specific ability to switch between phenotypic traits (migratory, proliferative or quiescent), we found that increasing EGF-receptor density leads to an earlier onset of switching events in the tumor system from the quiescent (**Q**) to migratory (**M**) and/or proliferative (**P**) cell phenotypes (**Fig. 4a**). Moreover, the cell line with the highest EGFR density (per cell, *iii*) switches earlier between proliferative and migratory traits (**Fig. 4b**), and vice versa (**Fig. 4c**).

**Figure 4a**

**Figure 4b**

**Figure 4c**

Thus, the larger the number of EGF receptors per cell, the earlier is the onset of phenotypic switching events mediated through the EGFR gene-protein interaction network.

In an attempt to delve into the possible molecular basis of such switching behavior and its earlier onset, we plotted the average values of the major components of our network over time (**Fig. 5**).The plots show that the cell-average of PLCγ concentration changes most gradually in the cell type with the lowest EGFR density (*i*) and faster in the cell types with increased EGFR (*ii*) and (*iii*) (**Fig. 5a**). Interestingly, although TGFα does not directly influence the phenotype in our model, our result shows that early increases in extracellular TGFα concentration (0.4-0.6 nM) correlate with the more invasive, high-EGFR density cell line (**Fig. 5b**).

**Figure 5a**

**Figure 5b**

## 4. DISCUSSION and CONCLUSIONS

Results from numerous experimental and clinical studies support the notion that the epidermal growth factor receptor is critically involved in various aspects of brain tumorigenesis. Employing our previously developed multiscale agent-based brain tumor model, we therefore investigated in here how variations on the receptor level affect both the sub-cellular as well as the multi-cellular tumor dynamics. Specifically, as a first step we varied the number of EGF-receptors within our EGFR-gene-protein module and simulated its impact across the scales of interest.

In accordance with experimental findings reported by Berens et al. (1996), our results show that the rate of spatio-temporal tumor expansion accelerates with increasing EGFR density per cell (Lund-Johansen et al., 1990), but reaches a plateau in the high receptor densities beyond which no further acceleration can be achieved. This increase and saturation of the expansion rate also supports results from our previous works (Mansury and Deisboeck, 2003; Mansury and Deisboeck, 2004a; 2004b) in which we could for instance show that up to a certain point a more precise spatial search process increases the rate of expansion of the entire tumor. While that work already implied receptor-mediated chemotaxis, the EGFR receptor, however, had not been explicitly modeled. Most importantly, our findings show now that early onset of phenotypic alternations is an intrinsic feature of such aggressive tumors. Interestingly, this refers to changes in both directions, i.e., from the proliferative to the migratory trait and vice versa. This means that *both* cellular phenotypes are necessary for overall rapid tumor expansion, which in turn hints on a critical role of the network's switching activity. Investigating this then further, we found that while cells with more EGFR molecules seem to take slightly longer to polarize, the resulting spatial patterns of PLCγ distribution are, however, qualitatively similar. Instead it appears that this earlier onset in 'switching' behavior is at least to some extent related to both, a faster change in average PLCγ concentration per cell and an early increase in TGFα protein secreted by these cells. Since, at the same time the gene expression level of TGFα and EGFR in the three cell lines considered does not differ, our results indicate that the EGFR dependent increased expansion rate depends largely on *post-translational* events. Interestingly, this switching between phenotypes could help explain the findings of immunostaining studies of glioma clinical samples, where EGFR and TGFα protein expression were found to correlate with increasing malignancy. While this did not correlate well with the Ki-67 mitotic index, it was however suggested to correlate with the early onset of autocrine signaling (von Bossanyi et al., 1998). From a therapeutic perspective, it is then noteworthy that EGFR kinase inhibitors were able to block this very same auto-stimulatory TGFα-EGFR loop in a subcutaneous glioma model in mice (El-Obeid et al., 2002). Furthermore, a PLCγ specific inhibitor that blocks binding and activation of growth factor receptors indeed reduced the infiltration rate of human glioblastoma spheroids (Khoshyomn et al., 1999).

As such, our finding may well be of interest for experimental and clinical cancer research alike as it indicates that changes in the EGFR pathway cannot be detected by monitoring gene expression data *alone*. Rather, proteomics data from tissue specimen should be included in the evaluation if predictive assessments of tumor growth dynamics are intended. We note that we did not yet examine in detail the impact of a concomitant EGFR gene-copy number increase on our molecular network, since data from human glioma samples are ambiguous with regards to the precise quantitative link between EGFR protein overexpression and gene-amplification (Fleming et al., 1992). While it is therefore a reasonable first step to vary the receptor protein amount per cell separately, we plan in future studies to investigate in more detail the initially stated close relationship between gene amplification and malignancy. Our preliminary findings using experimentally measured values of EGFR gene copy number increases in gliomas of up to 100-fold (Okada et al., 2003) indicate that increasing EGFR copies *in silico* do not linearly correlate with expansion velocity changes, thus seemingly further supporting the role of the protein in our admittedly limited network. While future work will also include implementing a more complex three-dimensional environmental setting, it is thus evident that one must strive to eventually anchor the EGFR pathway within a more extended sub-cellular interaction network, including for instance other tyrosine-kinase receptor cascades, apoptosis pathways and cell division pathways.

Yet, regardless of these current limitations, the presented results yield several experimentally testable hypotheses and thus support further refinement for and use of such multiscale models in interdisciplinary cancer research.

## ACKNOWLEDGEMENTS

This work has been supported in part by NIH grants CA 085139 and CA 113004 and by the Harvard-MIT (HST) Athinoula A. Martinos Center for Biomedical Imaging and the Department of Radiology at Massachusetts General Hospital.

## REFERENCES

Athale, C.A., Mansury, Y., Deisboeck, T.S., in press. Simulating the Impact of a Molecular 'Decision-Process' On Single Cell Phenotype and Multicellular Patterns in Brain Tumors. J. Theor. Biol. doi:10.1016/j.jtbi.2004.10.019.

Berens, M.E., Rief, M.D., Shapiro, J.R., Haskett, D., Giese, A., Joy, A., Coons, S.W., 1996. Proliferation and motility responses of primary and recurrent gliomas related to changes in epidermal growth factor receptor expression. J. Neurooncol. 27, 11-22.

Brightman, F., Fell, D., 2000. Differential feedback regulation of the MAPK cascade underlies the quantitative differences in EGF and NGF signalling in PC12 cells. FEBS Lett. 482, 169-174.

Chen, P., Xie, H., Sekar, M.C., Gupta, K., Wells, A., 1994. Epidermal growth factor receptor-mediated cell motility: phospholipase C activity is required, but mitogen-activated protein kinase activity is not sufficient for induced cell movement. J. Cell. Biol. 127, 847-857.

Chen, P., Xie, H., Wells, A., 1996. Mitogenic signaling from the egf receptor is attenuated by a phospholipase C-gamma/protein kinase C feedback mechanism. Mol. Biol. Cell. 7, 871-881.

de Jong, J.S., van Diest, P.J., van der Valk, P., Baak, J.P., 1998. Expression of growth factors, growth inhibiting factors, and their receptors in invasive breast cancer. I: An inventory in search of autocrine and paracrine loops. J Pathol. 184, 44-52.

Diedrich, U., Lucius, J., Baron, E., Behnke, J., Pabst, B., Zoll, B., 1995. Distribution of epidermal growth factor receptor gene amplification in brain tumours and correlation to prognosis. J Neurol. 242, 683-688.

Dittmar, T., Husemann, A., Schewe, Y., Nofer, J.R., Niggemann, B., Zanker, K.S., Brandt, B.H., 2002. Induction of cancer cell migration by epidermal growth factor is initiated by specific phosphorylation of tyrosine 1248 of c-erbB-2 receptor via EGFR. Faseb J. 16, 1823-1825. Epub 2002 Sep 1819.

Dowd, C.J., Cooney, C.L., Nugent, M.A., 1999. Heparan sulfate mediates bFGF transport through basement membrane by diffusion with rapid reversible binding. J. Biol. Chem. 274, 5236-5244.

El-Obeid, A., Hesselager, G., Westermark, B., Nister, M., 2002. TGF-alpha-driven tumor growth is inhibited by an EGF receptor tyrosine kinase inhibitor. Biochem. Biophys. Res. Commun. 290, 349-358.

Fleming, T.P., Saxena, A., Clark, W.C., Robertson, J.T., Oldfield, E.H., Aaronson, S.A., Ali, I.U. 1992 Amplification and/or overexpression of platelet-derived growth factor receptors and epidermal growth factor receptor in human glial tumors. Cancer Res. 52, 4550-4553.

Giese, A., Loo, M.A., Tran, N., Haskett, D., Coons, S.W., Berens, M.E., 1996. Dichotomy of astrocytoma migration and proliferation. Int. J. Cancer. 67, 275-282.

Khoshyomn, S., Penar, P.L., Rossi, J., Wells, A., Abramson, D.L., Bhushan, A., 1999. Inhibition of phospholipase C-gamma1 activation blocks glioma cell motility and invasion of fetal rat brain aggregates. Neurosurgery. 44, 568-577; discussion 577-568.

Kim, U.H., Kim, H.S., Rhee, S.G., 1990. Epidermal growth factor and platelet-derived growth factor promote translocation of phospholipase C-gamma from cytosol to membrane. FEBS Lett. 270, 33-36.

Knauer, D., Wiley, H., Cunningham, D., 1984. Relationship between epidermal growth factor receptor occupancy and mitogenic response. Quantitative analysis using a steady state model system. J. Biol. Chem. 259, 5623-5631.

Kues, T., Dickmanns, A., Luhrmann, R., Peters, R., Kubitscheck, U., 2001. High intranuclear mobility and dynamic clustering of the splicing factor U1 snRNP observed by single particle tracking. Proc Natl Acad Sci U S A. 98, 12021-12026. Epub 12001 Oct 12022.

Lauffenburger, D.A., Linderman, J. 1996. Receptors: Models for Binding, Trafficking, and Signaling. Oxford University Press, Oxford, U.K.

Libermann, T.A., Nusbaum, H.R., Razon, N., Kris, R., Lax, I., Soreq, H., Whittle, N., Waterfield, M.D., Ullrich, A., Schlessinger, J., 1985a. Amplification and overexpression of the EGF receptor gene in primary human glioblastomas. J Cell Sci Suppl. 3, 161-172.

Libermann, T.A., Nusbaum, H.R., Razon, N., Kris, R., Lax, I., Soreq, H., Whittle, N., Waterfield, M.D., Ullrich, A., Schlessinger, J., 1985b. Amplification, enhanced expression and possible rearrangement of EGF receptor gene in primary human brain tumours of glial origin. Nature. 313, 144-147.

Lund-Johansen, M., Bjerkvig, R., Humphrey, P.A., Bigner, S.H., Bigner, D.D., Laerum, O.D., 1990. Effect of epidermal growth factor on glioma cell growth, migration, and invasion in vitro. Cancer Res. 50, 6039-6044.

Lund-Johansen, M., Forsberg, K., Bjerkvig, R., Laerum, O.D., 1992. Effects of growth factors on a human glioma cell line during invasion into rat brain aggregates in culture. Acta Neuropathol (Berl). 84, 190-197.

Maly, I.V., Wiley, H.S., Lauffenburger, D.A., 2004. Self-organization of polarized cell signaling via autocrine circuits: computational model analysis. Biophys. J. 86, 10-22.

Mansury, Y., Deisboeck, T., 2003. The impact of "search precision" in an agent-based tumor model. J. Theor. Biol. 224, 325-337. doi:10.1016/S0022-5193(03)00169-3.

Mansury, Y., Deisboeck, T., 2004a. Simulating 'structure-function' patterns of malignant brain tumors. Physica A. 331, 219-232. doi:10.1016/j.physa.2003.09.013.

Mansury, Y., Deisboeck, T., 2004b. Simulating the time series of a selected gene expression profile in an agent-based tumor model. Physica D. 196, 193-204. doi:10.1016/j.physd.2004.04.008.

Mansury, Y., Kimura, M., Lobo, J., Deisboeck, T., 2002. Emerging patterns in tumor systems: simulating the dynamics of multicellular clusters with an agent-based spatial agglomeration model. J. Theor. Biol. 219, 343-370. doi:10.1006/jtbi.2002.3131.

Okada, Y., Hurwitz, E.E., Esposito, J.M., Brower, M.A., Nutt, C.L., Louis, D.N. 2003. Selection pressures of TP53 mutation and microenvironmental location influence epidermal growth factor receptor gene amplification in human glioblastomas. Cancer Res. 15, 413-416.

Owen, M.R., Sherratt, J.A., 1998. Mathematical modelling of juxtacrine cell signalling. Math. Biosci. 153, 125-150.

Owen, M.R., Sherratt, J.A., Wearing, H.J., 2000. Lateral induction by juxtacrine signaling is a new mechanism for pattern formation. Dev Biol. 217, 54-61. doi:10.1006/dbio.1999.9536.

Pfeuffer, J., Tkac, I., Gruetter, R., 2000. Extracellular-intracellular distribution of glucose and lactate in the rat brain assessed noninvasively by diffusion-weighted 1H nuclear magnetic resonance spectroscopy in vivo. J Cereb Blood Flow Metab. 20, 736-746.

Piccolo, E., Innominato, P.F., Mariggio, M.A., Maffucci, T., Iacobelli, S., Falasca, M., 2002. The mechanism involved in the regulation of phospholipase Cgamma1 activity in cell migration. Oncogene. 21, 6520-6529.

Schlegel, J., Merdes, A., Stumm, G., Albert, F.K., Forsting, M., Hynes, N., Kiessling, M., 1994. Amplification of the epidermal-growth-factor-receptor gene correlates with different growth behaviour in human glioblastoma. Int J Cancer. 56, 72-77.

Schoeberl, B., Eichler-Jonsson, C., Gilles, E.D., Muller, G., 2002. Computational modeling of the dynamics of the MAP kinase cascade activated by surface and internalized EGF receptors. Nat. Biotechnol. 20, 370-375.

Shvartsman, S., Wiley, H., Deen, W., Lauffenburger, D., 2001. Spatial range of autocrine signaling: modeling and computational analysis. Biophys. J. 81, 1854-1867.

Shvartsman, S.Y., Muratov, C.B., Lauffenburger, D.A., 2002. Modeling and computational analysis of EGF receptor-mediated cell communication in Drosophila oogenesis. Development. 129, 2577-2589.

Starbuck, C., Lauffenburger, D., 1992. Mathematical model for the effects of epidermal growth factor receptor trafficking dynamics on fibroblast proliferation responses. Biotechnol. Prog.. 8, 132-143.

von Bossanyi, P., Sallaba, J., Dietzmann, K., Warich-Kirches, M., Kirches, E. 1998. Correlation of TGF-alpha and EGF-receptor expression with proliferative activity in human astrocytic gliomas. Pathol Res Pract. 194, 141-147.

Wells, A., 1999. EGF receptor. Int. J. Biochem. Cell Biol. 31, 637-643.

Westermark, B., Magnusson, A., Heldin, C.H., 1982. Effect of epidermal growth factor on membrane motility and cell locomotion in cultures of human clonal glioma cells. J Neurosci Res. 8, 491-507.

Wiley, H., Cunningham, D., 1981. A steady state model for analyzing the cellular binding, internalization and degradation of polypeptide ligands. Cell. 25, 433-440.

# TABLE AND FIGURE CAPTIONS

**Table 1.** Depicted are the variables of the EGFR gene-protein interaction network model here, and the molecular species they represent. Where no references were available in the literature reasonable estimates were used instead.

**Figure 1.** **(a)** The molecular decision network consists of positive and negative feedbacks (*arrows* with + and –) and mass action kinetics (*black arrows*) between genes and proteins. The cellular decisions are indicated with *grey arrows*. **(b)** Each cell in the model consists of a central nuclear space surrounded by cytoplasm and a membrane compartment. These 2D compartments are further divided into four sub-compartments in the cardinal directions, each connected to two others. The *gray* sub-compartmental region **(c)** shows intra-compartmental flux determined by a rate of inflow ($\boldsymbol{k_{in}}$) and outflow ($\boldsymbol{k_{out}}$) as represented by *solid arrows*, while the *stippled arrows* indicate the exchange of mass between different compartments as a result of gene-protein interactions.

**Figure 2.** Plotted here is the time it takes for the first tumor cell to reach the edge of the glucose source (Y-axis) versus the number of EGFR receptors per cell (X-axis). The error bars result from 23 runs with different random number seeds. The points (*gray circles*) on the X-axes indicate increasing EGF-receptor counts of (*i*) $2.9 \times 10^4$, (*ii*) $1.5 \times 10^5$ and (*iii*) $1.59 \times 10^6$ per cell, as reported by (Lund-Johansen et al., 1990) for three different human glioma cell lines (*i*) D-263 MG, (*ii*) D-247 MG and (*iii*) D-37 MG, respectively.

**Figure 3.** Shown is the 2D cross-section of **(a)** a tumor spheroid system at the time point when the first cell invades the edge of the nutrient source (located in the NE quadrant), for the three different cell lines (see **Fig. 2**). **(b)** Depicted is the localization of PLCγ in this first cell

at five (rescaled) consecutive time points. The color bar indicates the PLCγ concentration in [nM].

**Figure 4.** **(a)** The number of events in which quiescent tumor cells (**Q**) switch their phenotype to a migratory or proliferative trait (**M/P**) (Y-axis) is plotted for the three cell lines (*i, ii* and *iii*) over time (X-axis). **(b)** Depicted are the phenotypic changes from proliferative (**P**) to migratory (**M**) and, in **(c),** from the migratory to the proliferative trait. The data is superimposed from 30 runs with different random number seeds.

**Figure 5.** Here the average concentration of **(a)** cellular PLCγ, and **(b)** neighboring extra-cellular TGFα (in [nM]) of all viable tumor cells is plotted on the Y-axis over time (X-axis) for the three cell lines *(i)*, *(ii)* and *(iii)*. The plots from multiple runs (n=30) with different random number seeds are superimposed on one another.

# FIGURES and TABLES

## Table 1.

| *Symbol* | *Variable* | *Initial Values [nM]* | *Reference* |
|---|---|---|---|
| $X_1$ | TGFα extracellular protein | 1 | (Dowd et al., 1999) |
| $X_2$ | EGFR cell surface receptor | 25 | (Maly et al., 2004) |
| $X_3$ | Dimeric TGFα-EGFR cell surface complex | 0 | - |
| $X_4$ | Phosphorylated active dimeric TGFα-EGFR cell surface complex | 0 | - |
| $X_5$ | Cytoplasmic inactive dimeric TGFα-EGFR complex | 0 | - |
| $X_6$ | Cytoplasmic EGFR protein | 0 | - |
| $X_7$ | Cytoplasmic TGFα protein | 1 | Estimate |
| $X_8$ | EGFR RNA | 1 | (Kues et al., 2001) |
| $X_9$ | TGFα RNA | 0 | (Kues et al., 2001) |
| $X_{10}$ | PLCγ Ca-bound | 1 | (Piccolo et al., 2002) |
| $X_{11}$ | PLCγ active, phosphorylated, Ca-bound | 1 | (Kim et al., 1990) |
| $X_{12}$ | Glucose cytoplasmic | 1 | (Pfeuffer et al., 2000) |
| $X_{13}$ | Glucose extracellular | 0 | - |

**Figure 1.**

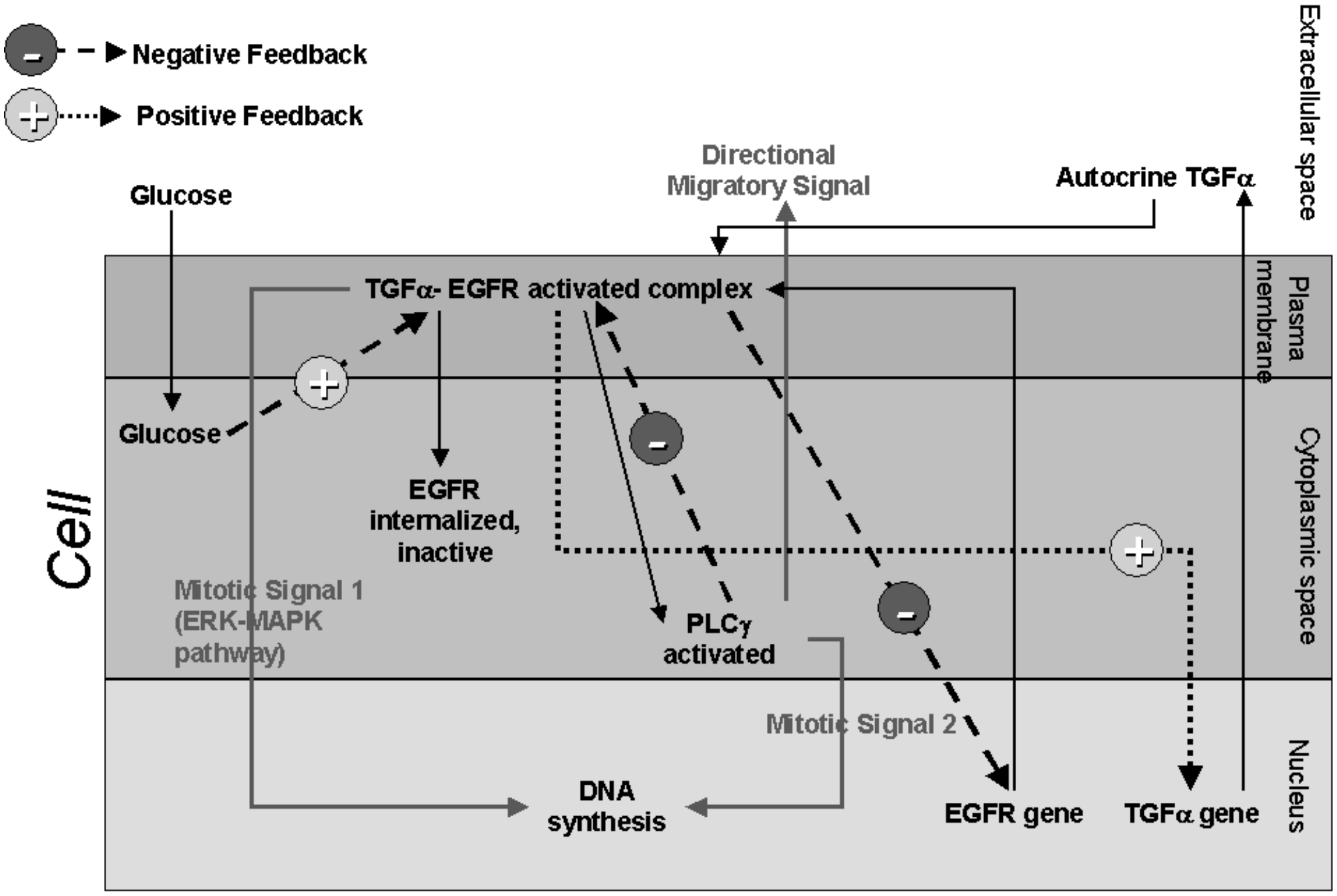

**(a)**

**(b)**

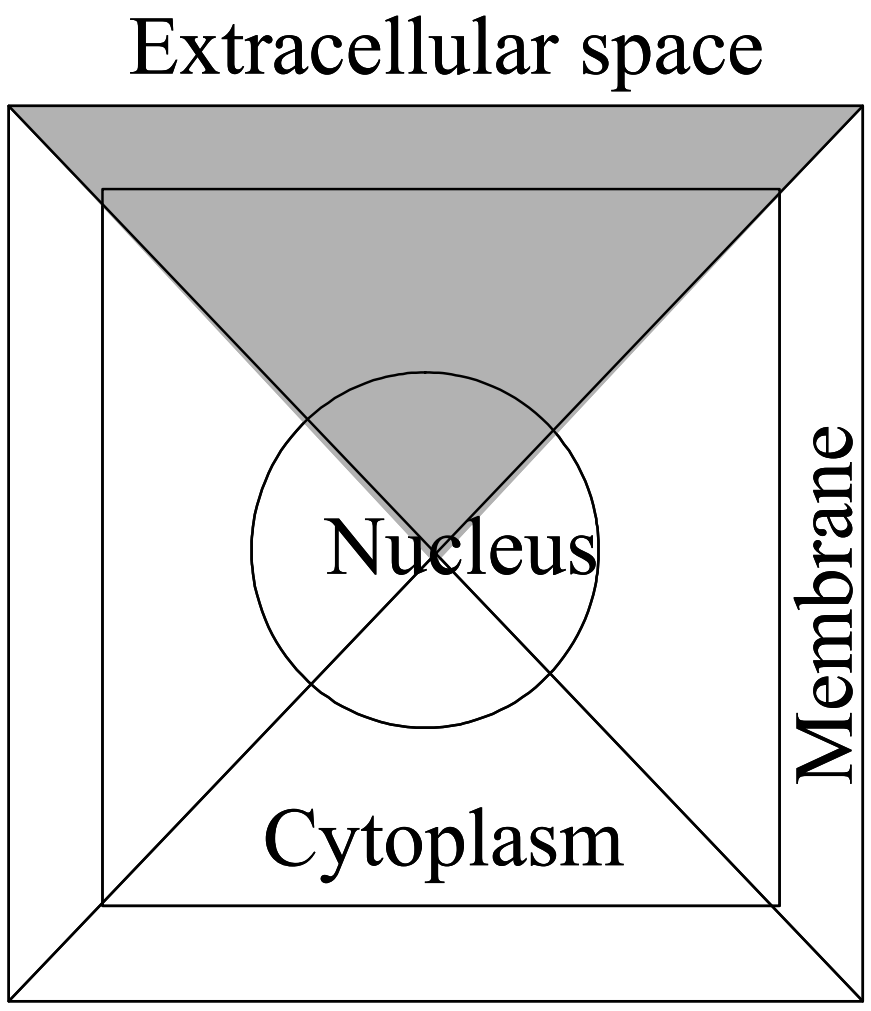
Extracellular space
Nucleus
Membrane
Cytoplasm

**(c)**

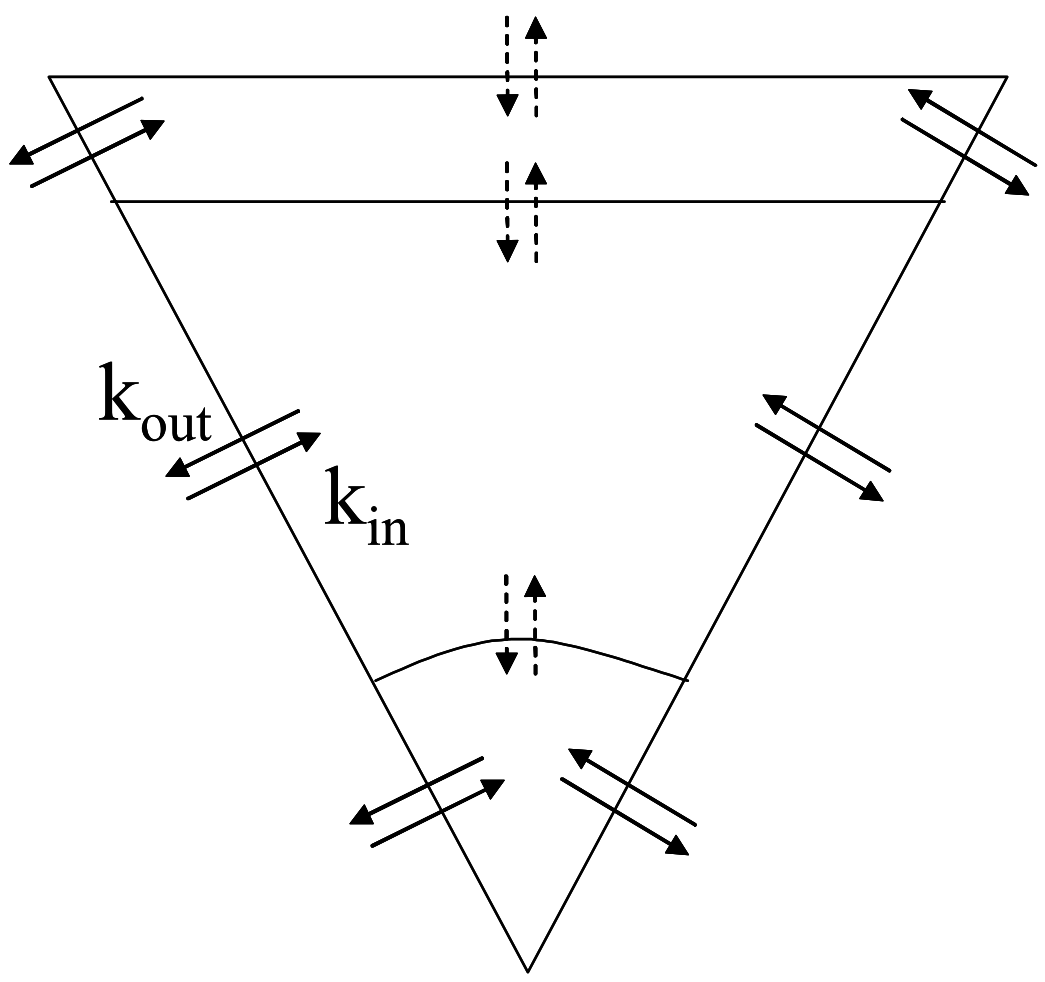
$k_{out}$
$k_{in}$

**Figure 2.**

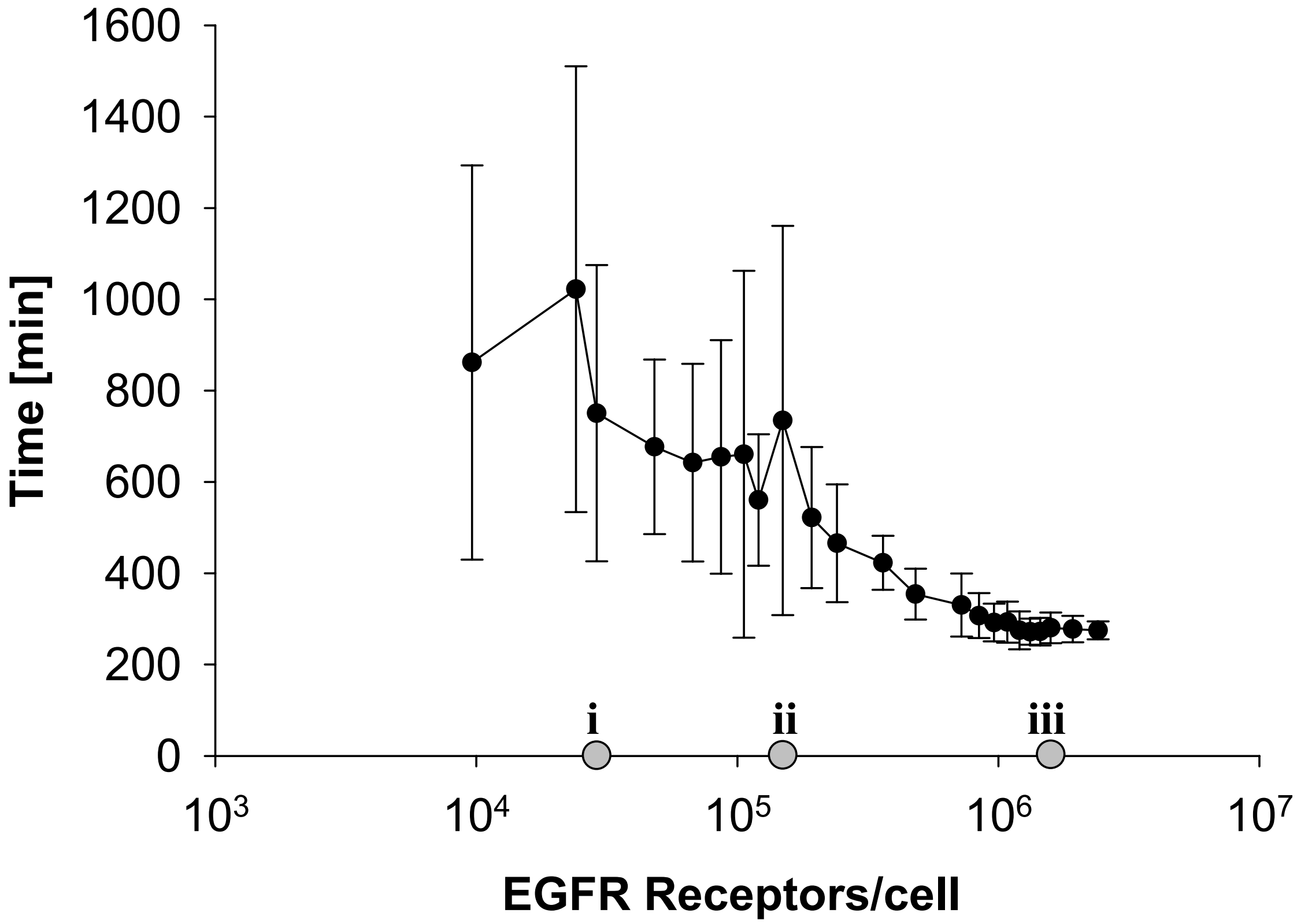

**Figure 3.**

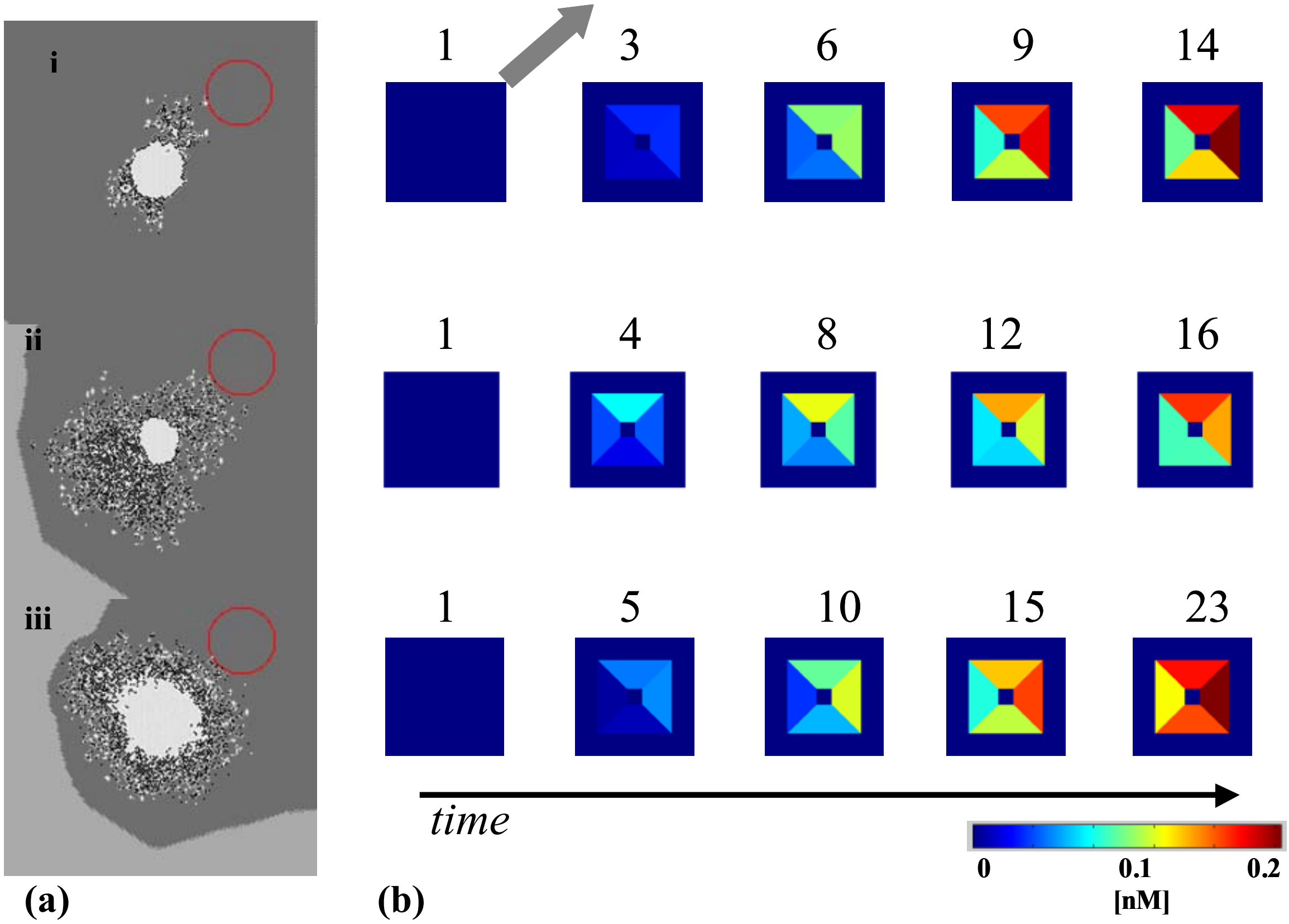

**(a)** **(b)**

**Figure 4.**

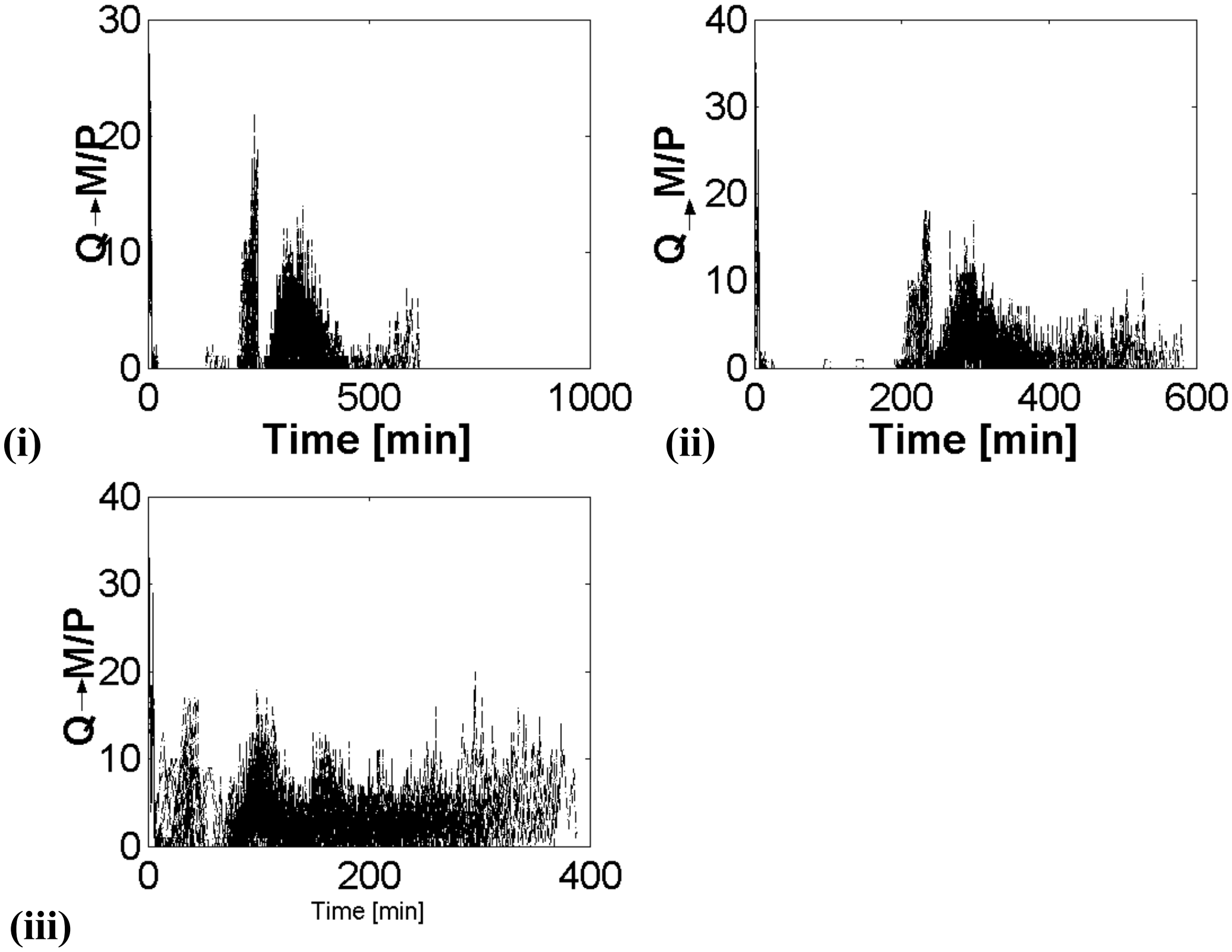

**(a)**

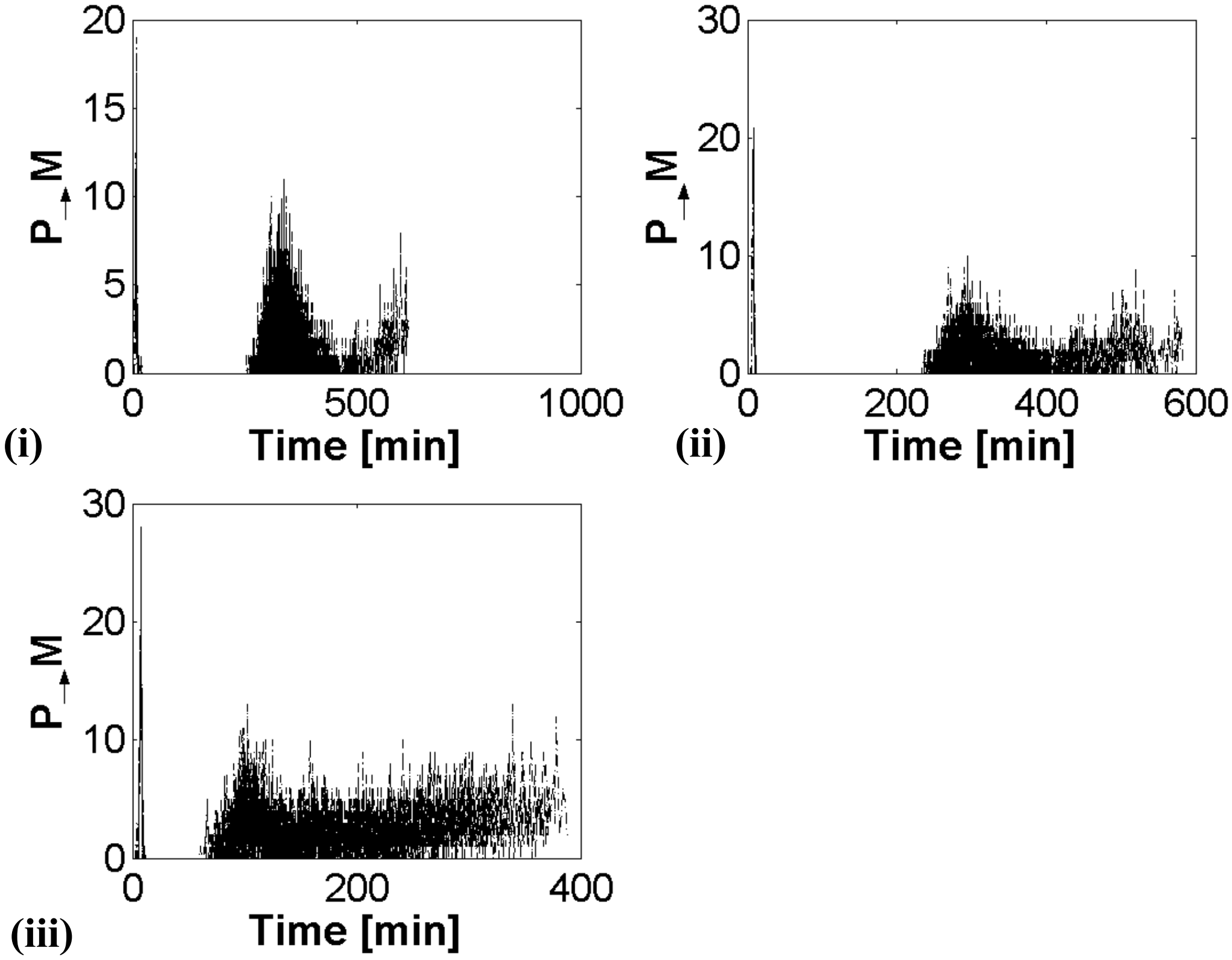

**(b)**

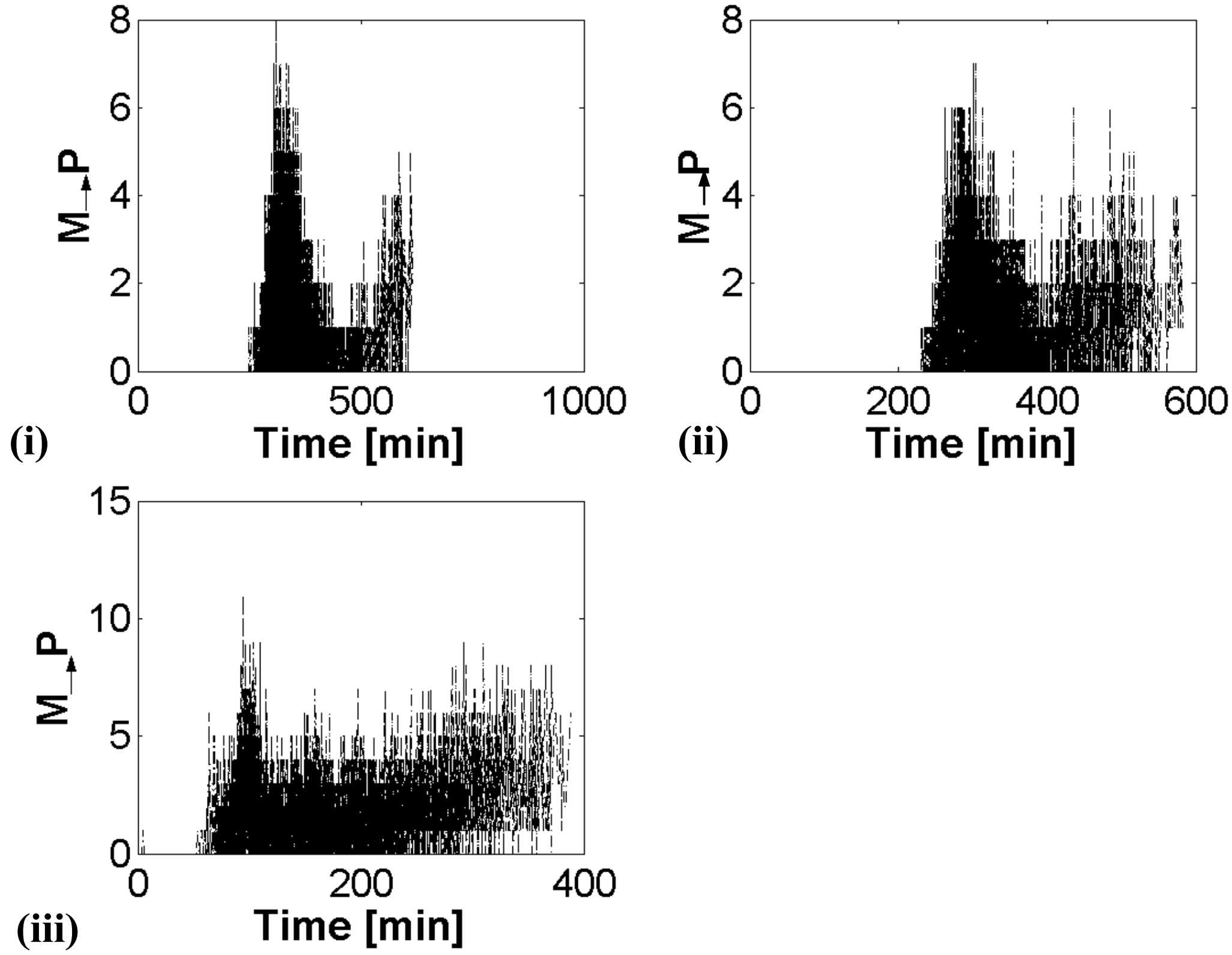

**(c)**

**Figure 5.**

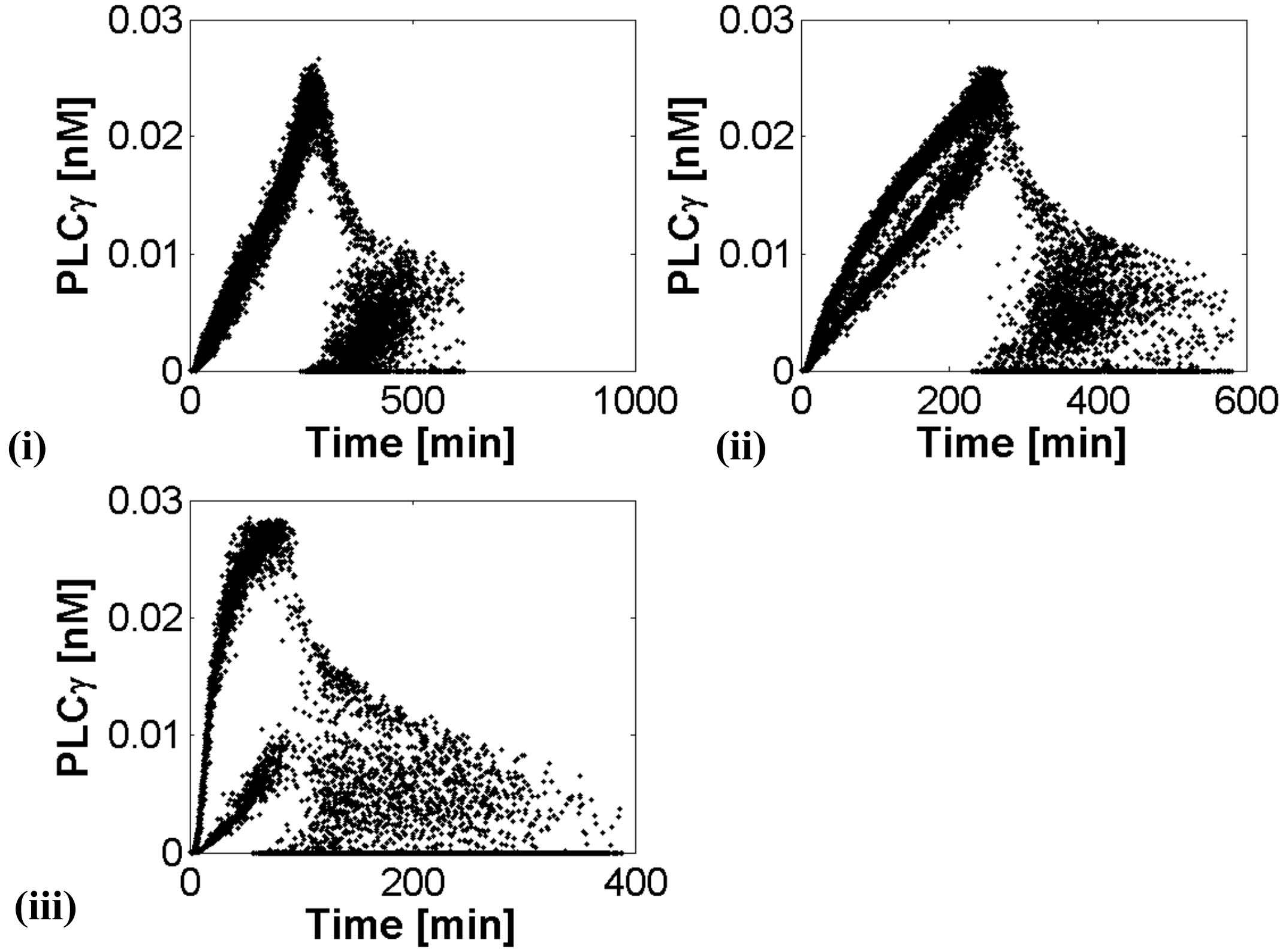

**(a)**

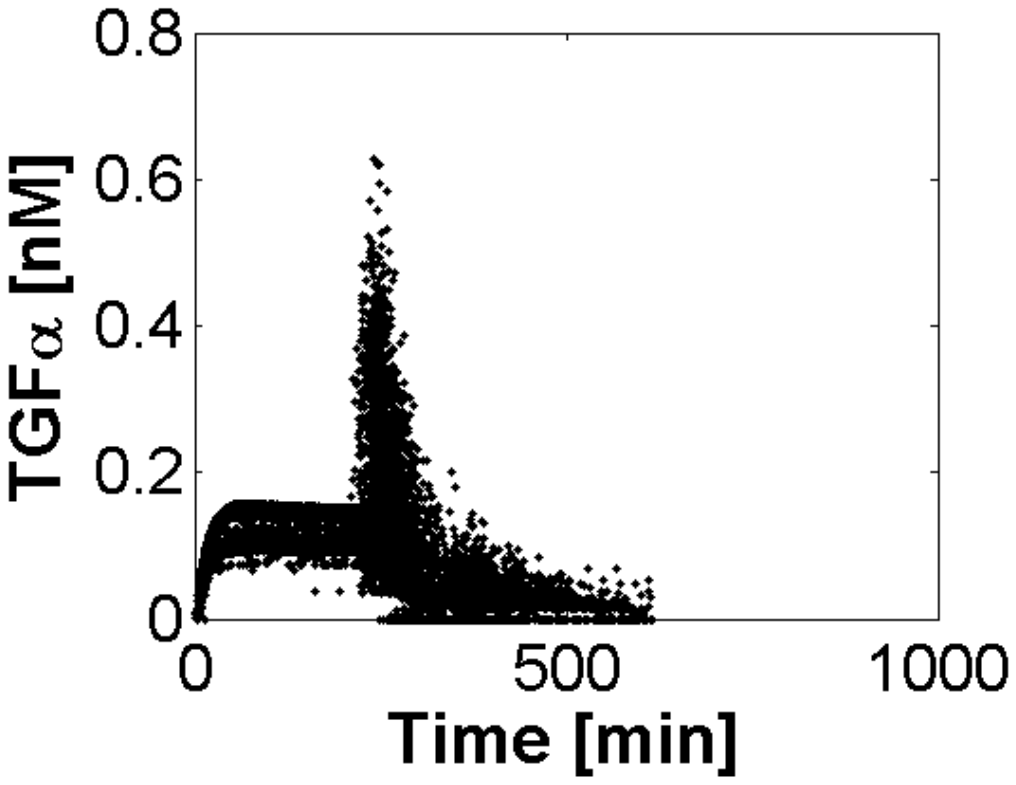

(i)

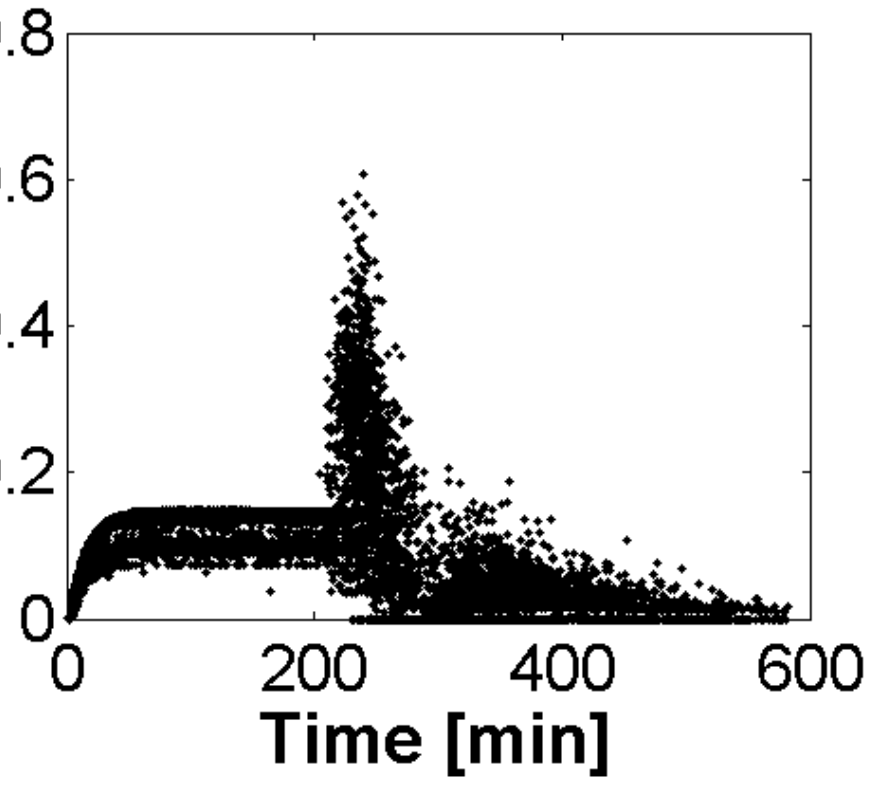

(ii)

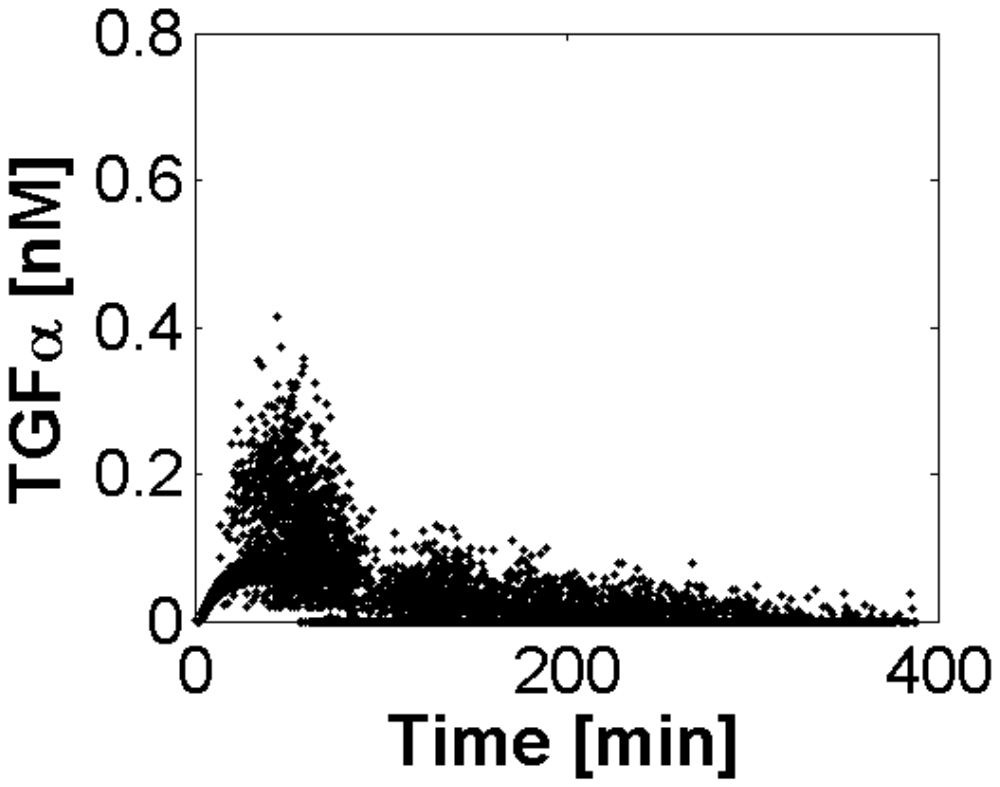

(iii)

(b)